\documentclass[a4paper]{jpconf}
\usepackage{graphicx}
\begin{document}

\title{Localized waves in the nonlinear rhombic waveguide array }

\author{A.I. Maimistov, E.I. Lyashko, E.O. Elyutin}

\address{
Department of Solid State Physics and Nanostructures, National
Nuclear Research University, Moscow Engineering Physics Institute,
Moscow, 115409 Russia }

\ead{aimaimistov@gmail.com,~ ostroukhova.ei@gmail.com }

\begin{abstract}
Solitary electromagnetic waves propagating along the waveguides
forming a rhombic one-dimensional lattice are considered. Two
waveguides that are part of the unit cell are assumed to be made of
an optical linear material, while the third waveguide from the same
array is composed of material with the cubic nonlinearity. The
equations of the coupled waves spreading in each waveguide are
solved under some approximation. These solutions represent the
breather like solitary waves, which are akin to three component
soliton.
\end{abstract}

\section{Introduction}
The first demonstration of the discrete photonic device was
presented in \cite{Somekh:73}. It was shown that the array of a
closely spaced waveguides acts as the Bragg grating. In
\cite{Pertsch:02,Staron:05,Ropke:11} the anomalous refraction and
diffraction in a discrete photonic system produced from an array of
coupled waveguides was experimentally studied. In waveguide arrays
with linearly varying propagation constants the optical Bloch
oscillations observed \cite{Pesch:98} The existence of localized
modes with equidistant wave-number spacing that do not undergo
diffraction is analytically proved \cite{Pesch:98,Szame:07}.

Over the last ten years the periodic one-dimensional or
two-dimensional lattices of waveguides whose coupling is due to
disturbed total internal reflection are the popular models of the
discrete photonics \cite{Longhi:09,Szameit:10,Maimi:16}.
\begin{figure}[h]
\includegraphics[width=15pc]{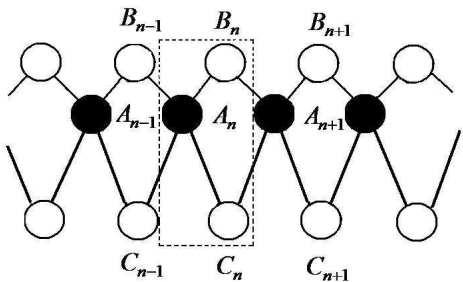}\hspace{2pc}%
\begin{minipage}[b]{14pc}
\caption{\label{Figa1} Binary rhombic array of waveguides. The unit
cell of a one-dimensional array is marked with a rectangle. }
\end{minipage}
\end{figure}

If the unit cell of the waveguide lattice contains more than two
"atoms"\, the photon spectrum has several branches. There are
conditions where one or more branches have zero curvature in a space
of quasi-pulses. The expression for the light wave frequency does
not depend on the transverse wave numbers. The corresponding
spectral bands are called flat bands. The fields attributed to a
superposition of modes from the flat band remain localized on the
waveguide array.  In other words, theses fields correspond to beams,
which are free of discrete diffraction.

A rhombic waveguide array (Fig.\ref{Figa1}) is the example of the
discrete medium where the photon spectrum has one flat band and two
usual bands \cite{Flach:14,Longhi:14a}. In the linear case existence
of the localized flat band modes in the rhombic waveguide array was
experimentally demonstrate \cite{Mukher:15,Mukher:17}. Recent review
devoted to the optical system with a flat band is \cite{Leykam:18}.
In general case in a nonlinear rhombic waveguide array the modes of
the all band are stirred due to waveguide material nonlinearity. The
diffraction is restored \cite{Zegadlo:17,Maim:17}.

In this paper the binary nonlinear rhombic waveguide array will be
considered. Unit cell contains one nonlinear waveguide and two
linear ones. In \cite{Maim:18} the waves localized on the axis of
waveguides was considered. Oppositely, here the waves will be
assumed localized on transverse direction but along the waveguide
direction waves are not localized.

\section{Basic equations}

Let the $A_n$, $B_n$ and $C_n$ be  the normalized field amplitudes
in a waveguide of the $n$th unit cell (Fig.\ref{Figa1}). Evolution
of theses values is governed by the following system of equations
\cite{Maim:18}
\begin{eqnarray}
  && i(\partial_{\tau}+ \partial_{\zeta})A_n + (B_n+B_{n-1}) +\gamma (C_n+C_{n-1}) +\mu|A_n|^2A_n =0, \label{eq:1} \\
  && i(\partial_{\tau}+ \partial_{\zeta})B_n + (A_n+A_{n+1}) =0, \label{eq:2} \\
  && i(\partial_{\tau}+ \partial_{\zeta})C_n + \gamma(A_n+A_{n+1}) =0, \label{eq:3}
\end{eqnarray}
where $\zeta$ is a dimensionless coordinate measured in units of
coupling length between the waveguides of types $\mathrm{A}$ and
$\mathrm{B}$, $\tau$ is dimensionless time. Here the symbol $
\partial_s$ is used instead of $\partial/\partial s$. Parameter
$\gamma$ is coupling constant ratio between waveguides of types
$\mathrm{A}$ and $\mathrm{C}$. Waveguides of types $\mathrm{B}$ and
$\mathrm{C}$ are made of an optical linear material, while the
waveguide of type $\mathrm{A}$ is composed of material with the
cubic nonlinearity. Parameter $\mu$ considers the nonlinear
properties of waveguides of type $\mathrm{A}$.

If one introduce new variables $F_n=B_n+\gamma C_n$ and
$G_n=B_n-\gamma C_n$, then the system of equations
(\ref{eq:1})--(\ref{eq:3}) will reads
\begin{eqnarray}
  && i(\partial_{\tau}+ \partial_{\zeta})A_n + (F_n+F_{n-1}) +\mu|A_n|^2A_n =0, \label{eq:4} \\
  && i(\partial_{\tau}+ \partial_{\zeta})F_n +(1+\gamma^2) (A_n+A_{n+1}) =0, \label{eq:5} \\
  && i(\partial_{\tau}+ \partial_{\zeta})G_n + (1- \gamma^2)(A_n+A_{n+1}) =0. \label{eq:6}
\end{eqnarray}

It should be remarked that $F_n$ and $A_n$ are coupled to one
another though equations (\ref{eq:4}) and (\ref{eq:5}). The
amplitudes $G_n$ are determined by only $A_n+A_{n+1}$.

\section{Continuum approximation }

Let us $A_n=(-1)^n\tilde{A}_n$ and $F_n=(-1)^n\tilde{F}_n$ where
$\tilde{A}_n$ and $\tilde{F}_n$ are the slowly varying variables as
the function of $n$. The second proposition is
$$ \tilde{A}_n(\tau,\zeta) =
a_n\exp[i\beta(\zeta+\tau)], \quad \tilde{F}_n(\tau,\zeta) =
f_n\exp[i\beta(\zeta+\tau)].
$$
Then the system of equations (\ref{eq:4}) and (\ref{eq:5}) reduces
to
\begin{eqnarray}
  &&-\beta a_n+
  (f_n-f_{n-1})+\mu|a_n|^2a_n =0, \label{eq:4d} \\
  && -\beta f_n+
  (1+\gamma^2)(a_n-a_{n+1})=0, \label{eq:5d}
\end{eqnarray}
These equations are not contain the imaginary unit, hence the
variables $a_n$ and $f_n$ are real ones. Elimination $f_n$ results
in the equation for $a_n$:
\begin{equation}\label{eq:7}
    \beta^2a_n+ (1+\gamma^2)(a_{n-1}+a_{n-1}-2 a_n) -\beta\mu
    a_n^3 =0.
\end{equation}

Now the continuum approximation will be done. Variable $a_n$ will be
considered as the function of $\xi = n\delta l$ ($\delta l$ is the
lattice parameter). If the $a_n$ varies only slightly over several
$\delta l$, then the following approximation
$$
a_{n\pm 1} \approx a \pm \frac{\partial a}{\partial\xi} +
\frac{1}{2} \frac{\partial^2 a}{\partial\xi^2},
$$ can be assumed. In this approximation equation
 (\ref{eq:7}) is reduced to
\begin{equation}\label{eq:c:7}
    \beta^2a + (1+\gamma^2)\frac{\partial^2 a}{\partial\xi^2} -\beta\mu
    a^3 =0.
\end{equation}
The amplitude $f$ is
$$
\beta f = -(1+\gamma^2)\frac{\partial a}{\partial\xi}.
$$

Solution of the equation (\ref{eq:c:7}) can be found by standard
way. Under condition that $a(\xi) \to 0$ at $|\xi|\to\infty$,
solution of the (\ref{eq:c:7}) takes the following form
\begin{equation}\label{eq:an:8}
a(\xi) = \sqrt{\frac{2\beta}{\mu}}\mathrm{sech}\left(\frac{\beta
\xi}{\sqrt{1+\gamma^2}}\right).
\end{equation}

\begin{figure}[h] \centering
\begin{minipage}{14pc}
\includegraphics[width=14pc]{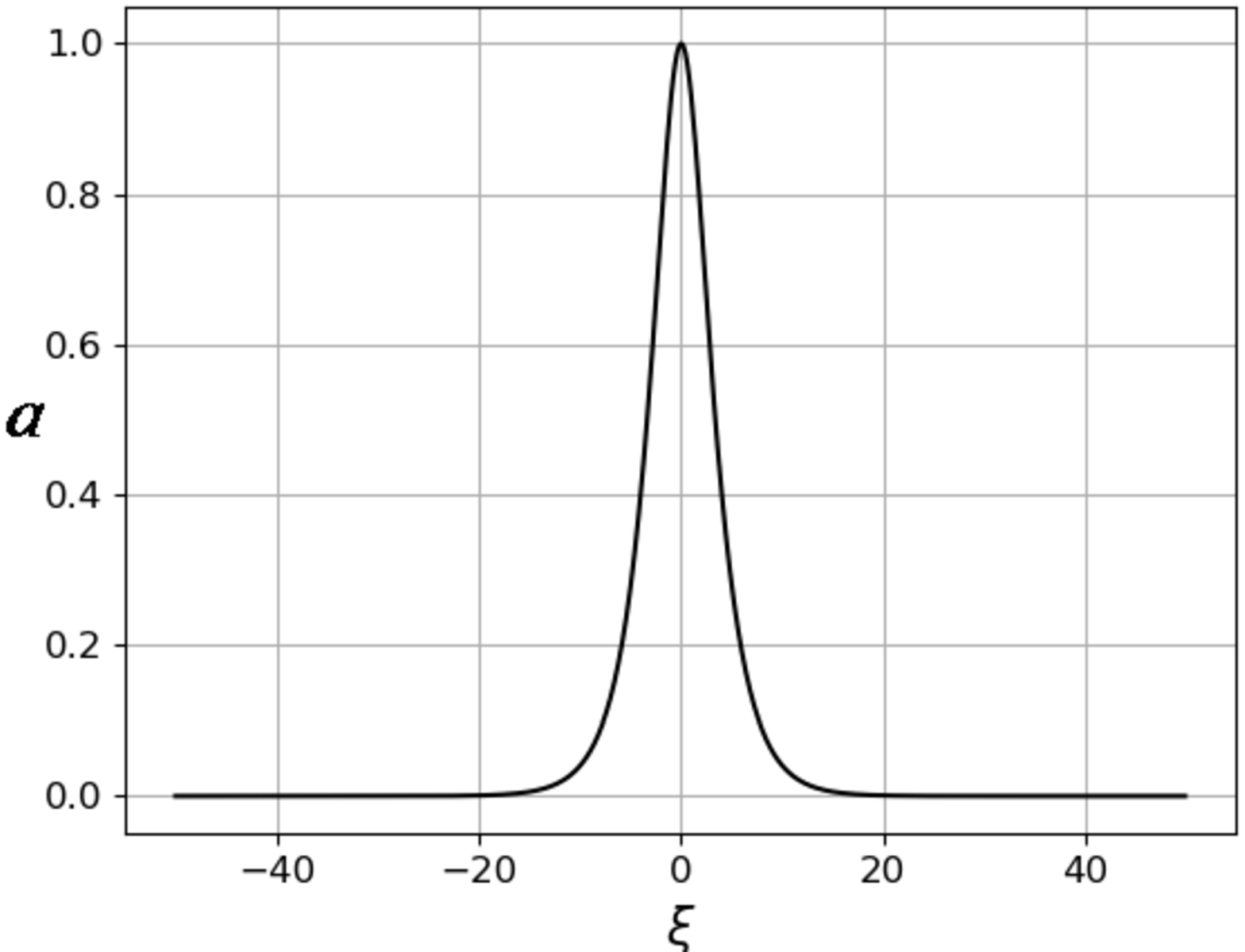}
\caption{\label{Fig:a}Slowly varying envelope  $a(\xi)$ for
localized wave in the waveguides of type $\mathrm{A}$.}
\end{minipage}\hspace{2pc}%
\begin{minipage}{14pc}
\includegraphics[width=14pc]{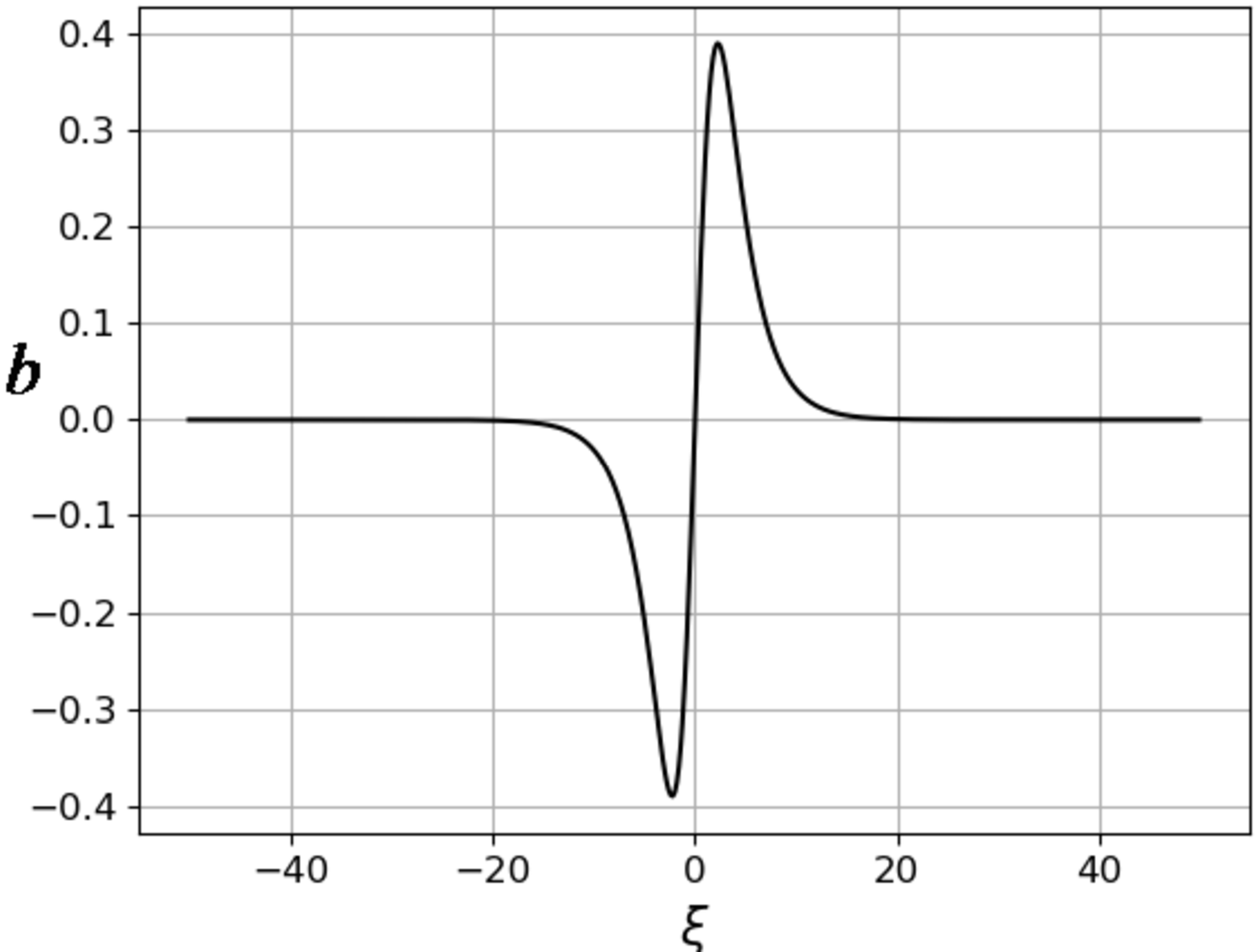}
\caption{\label{Fig:b} Slowly varying envelope $b(\xi)$ for
localized wave in the waveguides of type $\mathrm{B}$.}
\end{minipage}
\end{figure}

Taking into account the relationship between $G_n$, $F_n$, $B_n$ and
$A_n$ one can find the following expressions
\begin{eqnarray}
&&A_n(\tau,\zeta)=(-1)^n a(\xi)\exp[i\beta(\zeta+\tau)],\nonumber\\
&&B_n(\tau,\zeta)=(-1)^n b(\xi)\exp[i\beta(\zeta+\tau)],\label{eq:bn:10}\\
&&C_n(\tau,\zeta)=(-1)^n c(\xi)\exp[i\beta(\zeta+\tau)], \nonumber
\end{eqnarray}
where $\xi = n\delta l$, $a(\xi)$ is defined by equation
(\ref{eq:an:8}), $b(\xi)$ and $c(\xi)$ are equal to
\begin{eqnarray}
  && b(\xi)=\frac{1}{\sqrt{1+\gamma^2}} \sqrt{\frac{2\beta}{\mu}}\tanh\left(\frac{\beta
\xi}{\sqrt{1+\gamma^2}}\right)\mathrm{sech}\left(\frac{\beta
\xi}{\sqrt{1+\gamma^2}}\right),  \label{eq:bn:9} \\
  && c(\xi)= \frac{\gamma}{\sqrt{1+\gamma^2}} \sqrt{\frac{2\beta}{\mu}}\tanh\left(\frac{\beta
\xi}{\sqrt{1+\gamma^2}}\right)\mathrm{sech}\left(\frac{\beta
\xi}{\sqrt{1+\gamma^2}}\right).  \label{eq:cn:9}
\end{eqnarray}

Slowly varying envelopes of the electric field distributions
(\ref{eq:bn:10})  are presented on Fig.\ref{Fig:a} and
Fig.\ref{Fig:b}. These plots was calculated by using (\ref{eq:an:8})
and (\ref{eq:bn:9}) at $\beta = 0.5$, $\gamma = 0.8$ and $\mu =1$.
There only $a(\xi)$ and $b(\xi)$ are presented as the difference
between $c(\xi)$ and $b(\xi)$ is trivial.

\section{Conclusions}

The waves propagating along the axis of a periodic one-dimensional
binary rhombic array formed by waveguides of different types are
investigated. For the system of coupled mode equations approximate
solution was found in the case of the cubic nonlinearity of a
central waveguide line (Fig. 1). Two other lines were composed of
waveguides made of linear material. The obtained solution describes
a nonlinear solitary wave, which is akin to three component soliton.

It should be remarked that the electric fields in waveguides of
$\mathrm{A}$ and $\mathrm{B}$ types are identical in phase.
Localization of the fields is due to a dynamical reason. We not
consider the phenomenon of the flat band, where localization stems
from the interference of the electromagnetic fields
\cite{Leykam:18,Maim:Patrik:16,GMHF:Malomed:16}.

\section*{Acknowledgment}
The work was supported by the Russian Foundation for Basic Research
(Grant N 18-02-00278)


\end{document}